\def\spose#1{\hbox to 0pt{#1\hss}}

\def\multleft#1{\hbox to size{\vbox {\halign {\lft{##}\cr #1}}\hfill}\par}
\def\multright#1{\hbox to size{\vbox {\halign {\rt{##}\cr #1}}\hfill}\par}

\def\today{\ifcase\month\or January\or February\or March\or April\or May\or
      June\or July\or August\or September\or October\or November\or December\fi
      \space\number\day, \number\year}
\def\$<$ {\thinspace}
\def\s{\hbox{\phantom{5}}}	%one space
\def\cm{{\rm\thinspace cm}}

\def\erg{{\rm\thinspace erg}}

\def\g{{\rm\thinspace g}}
\def\K{{\rm\thinspace K}}
\def\keV{{\rm\thinspace keV}}
\def\km{{\rm\thinspace km}}
\def\kpc{{\rm\thinspace kpc}}

\def\Msun{\hbox{$\rm\thinspace M_{\odot}$}}
\def\pc{{\rm\thinspace pc}}

\def\s{{\rm\thinspace s}}
\def\yr{{\rm\thinspace yr}}

\def\ergpcmsqps{\hbox{$\erg\cm^{-2}\s^{-1}\,$}}

\def\ergps{\hbox{$\erg\s^{-1}\,$}}

\def\kmps{\hbox{$\km\s^{-1}\,$}}

\def\pcm{\hbox{$\cm^{-3}\,$}}
\def\pcmsq{\hbox{$\cm^{-2}\,$}}
\def\pcmK{\hbox{$\cm^{-3}\K$}}

\def\H2{\hbox{H$_{2}$}}

\def\approxlt{\mathrel{\spose{\lower 3pt\hbox{$\sim$}}
        \raise 2.0pt\hbox{$<$}}}
\def\approxgt{\mathrel{\spose{\lower 3pt\hbox{$\sim$}}
        \raise 2.0pt\hbox{$>$}}}

\documentstyle[psfig]{mn}
\include{defn}

\begin{document}
\hsize=6truein

\title[Exciting molecular hydrogen in the central galaxies of cooling flows]{Exciting molecular hydrogen in the central galaxies of cooling flows}
\author[R.J.~Wilman et al.]
{\parbox[]{6.in} {R.J.~Wilman$^{1,2}$, A.C.~Edge$^3$, R.M.~Johnstone$^1$, A.C.~Fabian$^{1}$, S.W.~Allen$^1$, and C.S.~Crawford$^1$ \\ \\
\footnotesize
1. Institute of Astronomy, Madingley Road, Cambridge CB3 0HA. \\
2. Leiden Observatory, PO Box 9513, 2300 RA Leiden, The Netherlands. \\
3. Department of Physics, University of Durham, Durham, DH1 7LE. \\ }}
\maketitle

\begin{abstract}
The origin of rovibrational \H2 emission in the central galaxies of cooling flow clusters is poorly understood. Here we address this issue using data from our near-infrared spectroscopic survey of 32 of the most line-luminous such systems, presented in the companion paper by Edge et al.~(2002).

We consider excitation by X-rays from the surrounding intracluster medium (ICM), UV radiation from young stars, and shocks. The v=1-0 K-band lines with upper levels within $10^{4}$\K~of the ground state appear to be mostly thermalised (implying gas densities $\approxgt 10^{5}$\pcm), with the excitation temperature typically exceeding 2000\K, as found earlier by Jaffe, Bremer \& van der Werf~(2001). Together with the lack of strong v=2-0 lines in the H-band, this rules out UV radiative fluorescence. 

Using the CLOUDY photoionisation code, we deduce that the \H2 lines can originate in a population of dense clouds, exposed to the same hot ($T \sim 50000$\K) stellar continuum as the lower density gas which produces the bulk of the forbidden optical line emission in the H$\alpha$-luminous systems. This dense gas may be in the form of self-gravitating clouds deposited directly by the cooling flow, or may instead be produced in the high-pressure zones behind strong shocks. Furthermore, the shocked gas is likely to be gravitationally unstable, so collisions between the larger clouds may lead to the formation of globular clusters.
\end{abstract}


\section{INTRODUCTION}
In the companion paper by Edge et al.~(2002) (hereafter paper 1) we presented H and K-band spectra of 32 of the most H$\alpha$ line-luminous cooling flow (CF) central cluster galaxies (CCGs) in the {\em ROSAT} Brightest Cluster Sample (Crawford et al.~1999). One or more of the rovibrational \H2 v=1-0 S(0)--S(7) lines were detected in 23 systems; one or both of [FeII]$\lambda\lambda$1.258,1.644 were seen in 14 systems, and there is also evidence for higher excitation \H2 lines (e.g. v=2-1 S(1), S(3)) and some coronal lines ([Si VI], [Si XI], [S XI] and [Ca VIII]). This sample quadruples the number of CF CCGs with \H2 detections, and builds on work by Jaffe \& Bremer~(1997), Falcke et al.~(1998), Donahue et al.~(2000) and Jaffe, Bremer \& van der Werf~(2001). Now that warm ($\sim 1000-2500$\K) molecular hydrogen is known to accompany ionized material in the core of a flow, our aim in this paper is to shed light on the associated excitation mechanisms.

Jaffe \& Bremer~(1997) found that the \H2 emission in CF CCGs is too luminous to be simply due to material passing through $\sim 2000$\K -- the temperature at which it is collisionally excited -- whilst cooling from $\sim 10^{7}$\K~(the well-known `H rec' problem). Based on a high v=1-0 S(1)/H$\alpha$ ratio, they implicated collisional excitation by suprathermal electrons liberated by X-ray photoionization deep within the cold, molecular core of the line-emitting clouds; the X-rays were hypothesized to originate in the surrounding intracluster medium (ICM). A similar model was advanced by Wilman et al.~(2000) for Cygnus A, which exhibits \H2, H recombination and [FeII] emission out to radii of 5\kpc. Flux ratios involving the v=1-0 S(1), S(3) and S(5) lines deviate markedly from the predictions of LTE, suggesting that non-thermal excitation is important. If the latter is provided by hard X-rays from the obscured quasar nucleus, the implied \H2 mass within 5\kpc~of the nucleus is $\sim 10^{10}$\Msun, accounting for $10^{9}$\yr~of mass deposition from the CF. 

From their HST NICMOS narrow-band imaging of the K-band \H2 emission in three CF CCGs, Donahue et al. deduced that the emission in NGC 1275 is principally due to the AGN (as did Krabbe at al.~2000), but that the more spatially-extended emission in A2597 and PKS 0745-191 is probably due to UV radiation from young stars. They based this conclusion on the close morphological correspondence between the \H2 and optical line emission, and on a comparison of the \H2/H$\alpha$ ratios with predictions for shock models, X-ray heating by the ICM, and UV fluorescence by stars and AGN. 

Most recently, Jaffe, Bremer \& van der Werf~(2001) presented UKIRT CGS4 K-band spectra of 7 well-known cooling flows. They observed that the relative strengths of the v=1-0 lines agree well with the predictions of LTE, but that there are deviations amongst the higher vibrational states. The thermal nature of the line production implies that the molecular gas is overpressurised by 2 to 3 orders of magnitude with respect to the X-ray and optical line emitting components. They sought to explain this using an ablative or `rocket' model of the gas system in which X- or UV-radiation heats the surface layers of a cold, gravitationally bound molecular cloud to 2000\K, at which temperature the material ceases to be gravitationally bound and expands into space, ionizes, and emits the optical lines. They deduced ionized to molecular line ratios which are lower than those of starburst regions, indicating that alternative heating mechanisms are necessary. 

Of relevance to the analysis presented in this paper are the results of the CO survey for cold molecular gas in CF CCGs presented by Edge (2001), many of whose targets are also in our UKIRT survey. He detected CO emission in 16 of these CFs, consistent with $10^{9-11.5}$\Msun~of molecular gas at $\sim 40$\K~(assuming a standard CO:\H2 conversion factor). The implied mass of cool \H2 correlates better with the H$\alpha$ luminosity than with the global X-ray mass deposition rate of the CF, suggesting that young stars are warming a population of molecular clouds. Since the physical conditions in the material probed by the K-band \H2 lines are likely to occupy an intermediate regime between the optical and CO line-emitting components, they offer the potential for further investigation into the link between the CO and H$\alpha$ emission. Indeed, we showed in paper 1 that there is a good correlation between the \H2~mass inferred from the CO emission and the 1-0~S(1) line luminosity, with the exceptions of the strong radio sources Cygnus A and PKS~0745-191.

\section{ANALYSIS}
We address the \H2 excitation mechanism from several angles: firstly, by considering the relative strengths of the various \H2 lines; secondly, by comparing the fluxes in the \H2 and H recombination lines and using established results for the origin of the optical line emission in these systems. Throughout we draw upon our own and published calculations for the \H2 emission under various physical conditions (viz. irradiation by UV and X-rays, and in shocks), as well as empirical results for the \H2 emission in Seyfert and starburst galaxies. A comparison between the widths of the \H2 and CO lines (the latter are determined to an accuracy of 50\kmps, and generally lie in the range 100--300\kmps~FWHM; Edge 2001) would also elucidate the link between the CO and \H2 emission, but is not possible due to the low resolution of our UKIRT spectra ($\sim 600$\kmps~FWHM at the centre of the K-band).

\subsection{\H2 line ratios}
\begin{figure*}
\psfig{figure=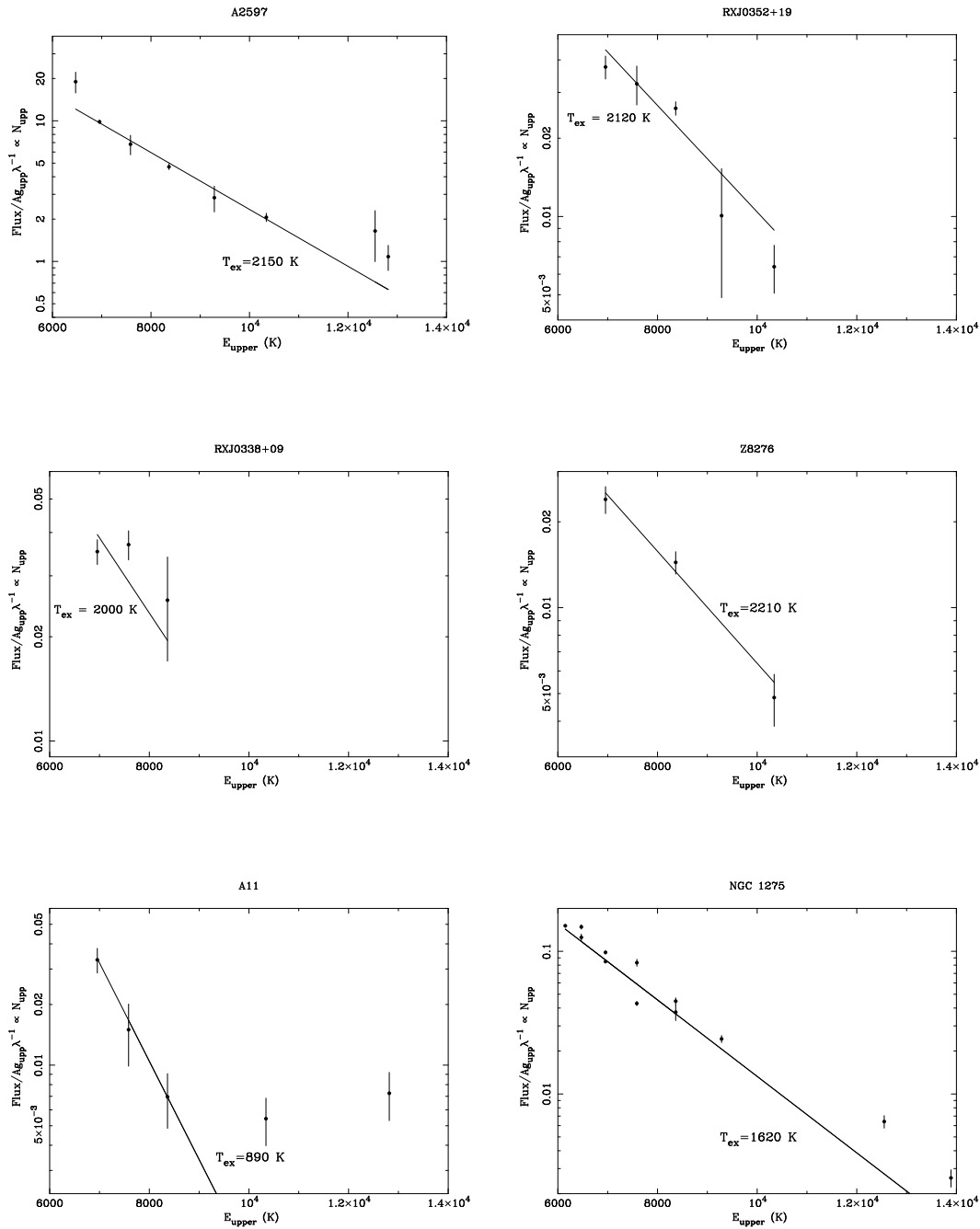,width=0.9\textwidth}
\caption{\normalsize LTE diagnostic plots for objects in the September 1999 dataset with three or more \H2 line detections. }
\label{fig:sept99lte}
\end{figure*}

\begin{figure*}
\psfig{figure=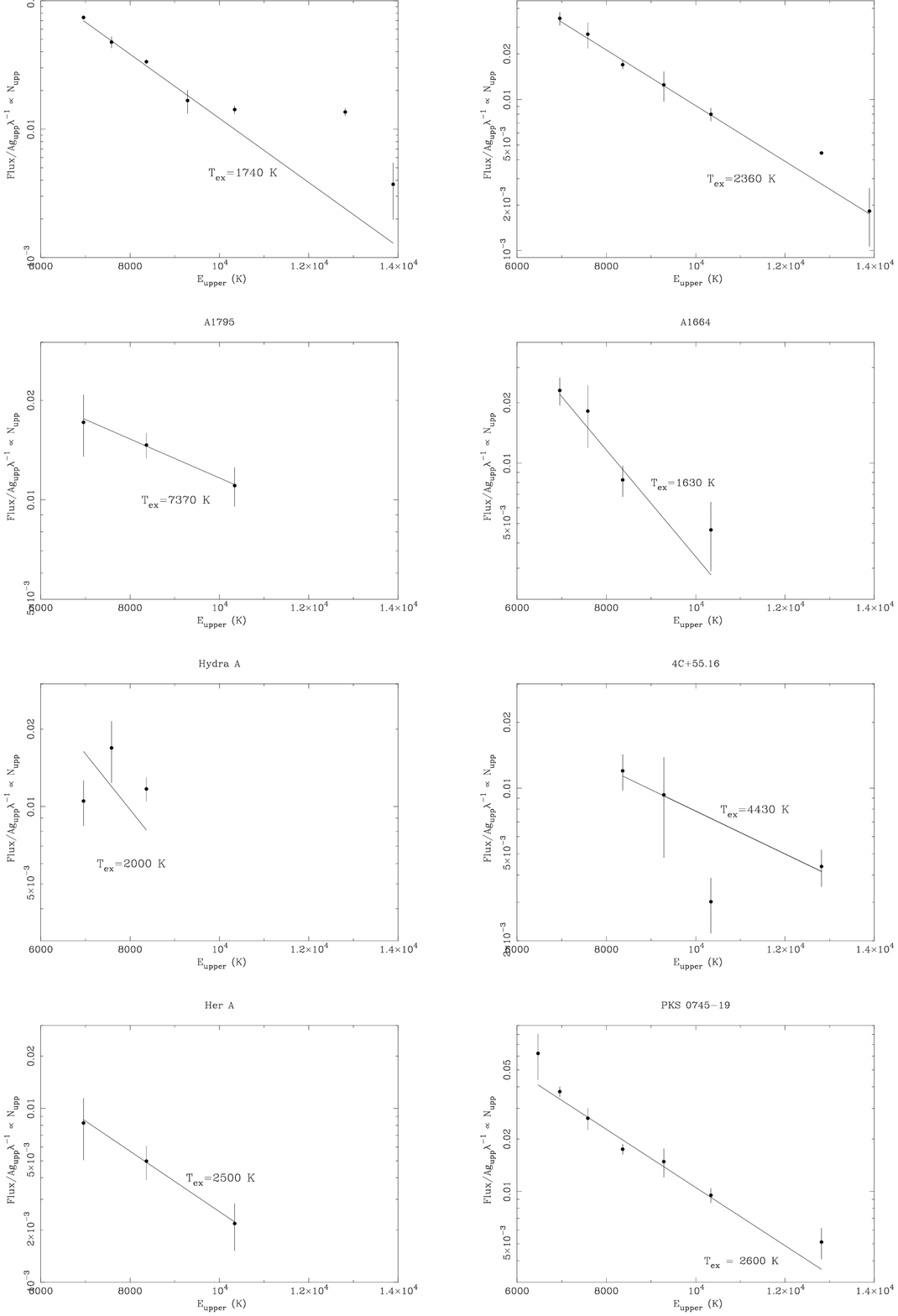,width=0.9\textwidth}
\caption{\normalsize As in Fig~\ref{fig:sept99lte} but for the March 2000 dataset. }
\label{fig:mar00lte}
\end{figure*}

We begin by comparing the observed \H2 line ratios with the predictions for thermal (collisional) excitation, which dominates in molecular gas with a density $n_{\rm{T}} \approxgt 10^{5}$\pcm, heated to a few thousand Kelvin by a source of X or UV radiation, or in shocks. Under these conditions, the occupation numbers of the excited rotation-vibration levels of the \H2 molecule will be in thermal equilibrium at a temperature $T_{\rm{ex}}$ equal to the kinetic temperature of the gas. Therefore, for a given object, the flux $\Im_{\rm{i}}$ in the $i$th emission line will satisfy the relation $log(\Im_{\rm{i}}\lambda_{\rm{i}}/A_{\rm{i}}g_{\rm{i}})= constant - T_{\rm{i}}/T_{\rm{ex}}$, where $\lambda_{\rm{i}}$ is the wavelength of the line, and $A_{\rm{i}}$, $T_{\rm{i}}$ and $g_{\rm{i}}$ are the spontaneous emission coefficient, excitation temperature and statistical weight of the upper level of the transition (the latter assumes an ortho:para \H2 abundance ratio of 3:1, appropriate where collisions dominate). LTE plots were constructed for 14 objects (those in which 3 or more \H2 lines were detected), and are shown in Figs.~\ref{fig:sept99lte} and \ref{fig:mar00lte} for the September 1999 and March 2000 objects, respectively. 

These figures show that where reliable measurement is possible, the v=1-0 lines with upper levels within $\sim 10^{4}$\K~of the ground state are thermally excited at temperatures in excess of 1600\K, with most being above 2000\K. The only exception to this is A11, where the excitation temperature for the v=1-0 S(1), S(2) and S(3) lines is $890 \pm 200$\K. Implied temperatures above 4000\K~are unphysical since \H2 is rapidly dissociated by collisions in hotter gas. Thus the result for A1795 should be regarded as spurious, especially since it conflicts with the results of Jaffe et al.~(2001) who found line ratios in LTE at approximately 2200\K. The discrepancy is probably due to our spectrum having a lower signal to noise ratio and the difficulty of assessing the true continuum underlying the v=1-0 S(3) and S(5) lines which fall in a region of low atmospheric transmission. The thermalised nature of the lower-lying transitions thus demands that the total atomic and molecular hydrogen density exceed $\sim10^{5}$\pcm, the critical density for collisional deexcitation. In combination with the measured temperatures, the implied pressures exceed those in the X-ray and optical line emitting systems by 2 to 3 orders of magnitude, and we therefore confirm the result of Jaffe et al.~(2001). Together with the absence of prominent v=2-0 lines in our H-band spectra, this also rules out UV radiative fluorescence in cold, low density gas. The latter mechanism was advocated by Donahue et al. for A2597 and PKS 0745-191 based on the value of the v=1-0~S(3)/H$\alpha$ ratio and energetic arguments. UV fluorescence may, however, play some role in the production of the higher excitation lines such as v=1-0~S(7) and v=2-1 S(1) and S(3) which are seen to deviate from LTE in Figs.~1 and 2. Collisional excitation by a suprathermal photoelectron (as modelled by Maloney, Hollenbach \& Tielens~1996) is an alternative non-thermal excitation mechanism which may also supply some of the flux in these lines.

Once in the thermal regime, it is difficult to determine the heat source for the gas from the \H2 lines alone. Mouri~(1994) has, however, shown that there is a demarcation in excitation temperature between heating by UV on the one hand, and by shocks or X-rays on the other; from the models of Sternberg \& Dalgarno~(1989) (hereafter SD89), he showed that UV radiation can heat the molecular gas to no higher than $1000$\K, whilst X-ray heating (from the models of Lepp \& McCray~1983) produces temperatures $\sim 2000$\K; from the models of Brand et al.~(1989), shocks occupy the upper end of the range in between. Thus, the temperatures derived from Figs.\ref{fig:sept99lte} and \ref{fig:mar00lte} appear to rule out UV heating and marginally favour X-ray heating over shocks. There are two caveats to this conclusion. Firstly, X-ray heating has been shown to be energetically incapable of powering the optical line emission in CF CCGs (see e.g. Crawford \& Fabian~1992). Secondly, many of the most luminous systems (with L(H$\alpha)>10^{41}$\ergps, i.e. most of the objects in our sample) have substantial populations of young stars, which produce enough UV radiation to power the observed H$\alpha$ emission (Crawford et al.~1999). The optical emission line spectra differ from those of `normal' HII regions, with strong Balmer lines relative to forbidden lines of [OIII], [NII] and [SII]; they can be reproduced with very hot stars ($T>40 000$\K) and a low mean ionisation parameter $U$ in the range --3.5 to --4.0 (Allen~1995).

%In contrast, Donahue et al. favoured some form of UV excitation for the production of the v=1-0~S(3) emission in A2597 and PKS 0745-191, but they arrived at this conclusion using just a single \H2/H$\alpha$ ratio and some energetic arguments. Furthermore, UV radiative fluorescence in cold, low density gas is also ruled out by the lack of prominent v=2-0 lines in our H-band spectra. The departure from LTE among some of the higher excitation lines in Figs.~\ref{fig:sept99lte} and \ref{fig:mar00lte}, such as v=1-0 S(7) and v=2-1 S(1) and S(3), suggests that non-thermal excitation also contributes. This could either be due to a collision with a suprathermal electron (as modelled by Maloney, Hollenbach \& Tielens~1996) or the absorption of a UV photon in the Lyman and Werner bands of the \H2 molecule. However, since such lines are typically weak, relatively unimportant contributors to the total \H2 emission, and the model parameter space large, we will not discuss them further.

\subsection{\H2 and HII line emission}
The HST NICMOS emission line imaging study of Donahue et al. revealed a close spatial correspondence between the \H2 and optical H$\alpha$ emission, suggesting a common excitation mechanism. 

\begin{figure}
\psfig{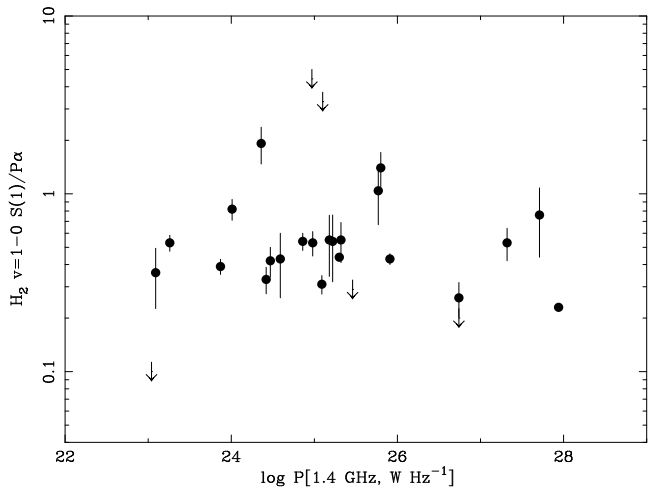}
\caption{\normalsize v=1-0 S(1)/Pa$\alpha$ versus monochromatic 1.4~GHz radio power.}
\label{fig:S1Palpha}
\end{figure}

We begin with an empirical comparison of the \H2 and HII line emission. Jaffe et al.~(2001) noted that in Seyfert, starburst and ultraluminous infrared galaxies (ULIRGs), the ratio v=1-0~S(1)/Br$\gamma \leq 1$. Assuming case B recombination this implies that v=1-0~S(1)/Pa$\alpha \leq 0.083$ for such objects. Fig.~\ref{fig:S1Palpha} shows, however, that for all our targets with detectable S(1), v=1-0~S(1)/Pa$\alpha > 0.2$. Thus, whatever the relationship between the \H2 and HII line emission in Seyferts, starburst galaxies and ULIRGs -- and this is by no means certain in itself -- it does not apply to our CF galaxies. Fig.~\ref{fig:S1Palpha} also shows no correlation with the power of the central radio source, which might have been expected if the associated AGN preferentially enhanced the production of the HII or \H2 lines. 

\begin{figure}
\psfig{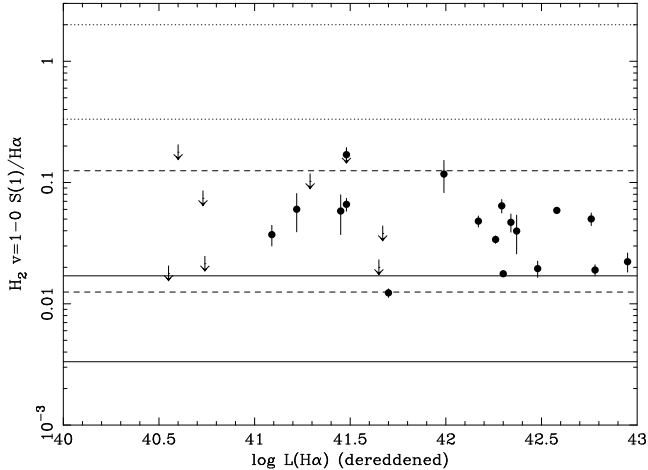}
\caption{\normalsize v=1-0 S(1)/H$\alpha$ versus H$\alpha$ luminosity where known (mostly taken from Crawford et al.~1999, corrected for intrinsic reddening which is typically $E(B-V) \sim 0.3$). The regions bounded by the pairs of solid and dotted lines correspond, according to Donahue et al., to UV fluorescence by an AGN and stellar sources, respectively. The dashed lines encompass the range for X-ray heating by the ICM. As discussed in section~2.3, however, {\em thermal} excitation by stellar UV radiation provides the most natural explanation for both the H$\alpha$ and \H2~emission.}
\label{fig:S1Halpha}
\end{figure}

Following Donahue et al. we also compare the observed v=1-0~S(1)/H$\alpha$ ratio with the predictions of various models, as shown in Fig.~\ref{fig:S1Halpha}. 
Since the H$\alpha$ and S(1) fluxes are both slit fluxes derived from separate observations at different position angles, this may introduce a scatter as large as a factor of 4 in v=1-0~S(1)/H$\alpha$. This is because the emission line images in Donahue et al. show that as much as half of the \H2~emission (and more of the H$\alpha$) arises in filamentary structures at radii greater than 1 arcsecond. 

 For UV excitation the v=1-0~S(1)/H$\alpha$ ratio is sensitive to the shape of the exciting spectrum near the Lyman edge: photons with $\lambda < 912$\AA~produce H$\alpha$, whilst those with $912 \leq \lambda \leq 1108 $\AA~(i.e. in the Lyman and Werner bands) lead to \H2 line emission. For a power-law spectrum typical of an AGN, they infer from the results of Black \& van Dishoeck~(1987) that the S(1)/H$\alpha$ ratio lies in the range 0.0033--0.017. Introducing a break by a factor 100 at the Lyman edge, to mimic a UV radiation field dominated by stars, yields a higher ratio of S(1)/H$\alpha \sim 0.33 - 2$. It should be noted, however, that the Black \& van Dishoeck models do not include collisional excitation processes for the vibrationally excited states and are thus not applicable to densities above $10^{5}$\pcm.

Donahue et al. also used the models of Maloney, Hollenbach \& Tielens~(1996) to compute the S(1)/H$\alpha$ ratio expected for X-ray heating by the ICM. For a thermal bremsstralung spectrum with $T=8$\keV~and a constant gas pressure of $10^{8}$\pcmK, S(1)/H$\alpha$ varies from 0.0125 to 0.125 as the total column density of the cloud is increased from $5 \times 10^{21}$ to $1.5 \times 10^{22}$\pcmsq. Since hard X-rays can excite \H2 emission through collision with suprathermal photoelectrons, the S(1)/H$\alpha$ can be increased further by augmenting the column density. Uncertainties in the shape of the UV/soft X-ray spectrum can also affect the contribution from the highly ionised layer at the cloud surface. Not withstanding such uncertainties, Fig.~\ref{fig:S1Halpha} shows that of the mechanisms considered by Donahue et al. X-ray heating is the most viable for the bulk of the objects, as found in section 2.1. However, in the next subsection we present calculations for thermal excitation in dense gas heated by the UV radiation from young stars, and describe how it provides a more natural explanation for the \H2~emission in these systems.

\subsection{CLOUDY photoionisation modelling}
To model {\em thermal} excitation of the \H2 lines in dense gas heated by UV radiation (as opposed to the {\em non-thermal UV} fluorescence modelled by Black \& van Dishoeck), we use the photoionisation code CLOUDY (version 94.00; Ferland~1996). As mentioned in section 2.1, we adopt a Kurucz model atmosphere at $50000$\K, with the ionisation parameter $U$ in the range --3.5 to --4.0 when incident on gas with hydrogen number density $n(H)=200$\pcm, in order to reproduce the optical emission line spectrum (this density matches that indicated by the [SII]$\lambda\lambda6717,6731$ doublet, see e.g. Voit \& Donahue~1997 for A2597). Note that CLOUDY does not list the \H2 lines separately but instead gives the `sum of the fluxes of the \H2 lines near 2$\mu$m' (which we denote by $\Sigma H_{\rm{2}}$). Since v=1-0~S(1) is typically the strongest \H2 line in our spectra, we do not for the moment draw a distinction between $\Sigma H_{\rm{2}}$ and v=1-0~S(1) in comparing model with observation. In the models we neglect dust and any contribution from magnetic pressure. Later in this sub-section we briefly discuss the effects of dust on the models and estimate the actual value of the 1-0~S(1)/$\Sigma H_{\rm{2}}$ ratio.

The inferred pressure difference between the optical and infrared emission line components clearly means that they cannot come from different parts of a single cloud in pressure equilibrium with the X-ray cluster gas. Such clouds would in any case have $\Sigma H_{\rm{2}}$/H$\alpha$ ratios several orders of magnitude below the observed range shown in Fig.~4. We can also dismiss the possibility that the two components arise at different depths within the same self-gravitating clouds, with the optical lines being produced in the surface layers which are in pressure balance with the diffuse X-ray gas: assuming hydrostatic equilibrium, the masses of the individual clouds would have to exceed $5 \times 10^{9}R_{\rm{pc}}^{\rm{2}}$\Msun~in order to produce the requisite pressure difference between the two components ($R_{\rm{pc}}$ being the cloud radius in \pc).

A more likely possibility is that the \H2 lines originate in a separate population of much denser, self-gravitating, clouds. As an example, consider clouds with $n(H)=10^{5}$\pcm~exposed to the above stellar continuum with $U=-6.2$ (i.e. with the same incident flux as that seen by low density, constant pressure clouds with $n(H)=200$\pcm~and $U=-3.5$). The resulting emission line spectrum has $\Sigma H_{\rm{2}}$/H$\alpha=0.1$ and the optical forbidden lines from ionised species are collisionally suppressed viz. [NII]$\lambda6584$/H$\alpha=0.042$, [SII]$\lambda6717$/H$\alpha=0.033$, but [OI]$\lambda6300$/H$\alpha=0.12$. Furthermore, the kinetic temperature in the \H2-emitting gas is 1000--2000\K~and the pressure is several times $10^{8}$\pcmK, as required. 

We thus propose that the variations in the [NII]/H$\alpha$, [SII]/H$\alpha$ and [OIII]/H$\beta$ ratios with H$\alpha$ luminosity, as shown in Figs.~7 and 8 of Crawford et al.~(1999), and likewise the variation of $\Sigma H_{\rm{2}}$/H$\alpha$ shown in Fig.~4 above, can be reproduced by a mixture of three separate emission line components. The relative line intensities within each component are fixed, but the fraction of the total H$\alpha$ emission that each contributes varies with the H$\alpha$ luminosity. The three components are: 

(a) Some mechanism (e.g. mixing layers; Crawford \& Fabian~1992) which produces the emission line ratios characteristic of the `type I' systems with H$\alpha$ luminosities below $10^{41}$\ergps; A2199 and A2052 are good examples of such systems, having [NII]/H$\alpha \simeq 2.3$, [SII]/H$\alpha \simeq 0.7$, [OIII]/H$\beta \sim 1$ (but with a large scatter), and no appreciable \H2 emission. \\

%(We exclude from consideration the [NII] only emitters which probably more closely resemble the LINER activity seen in normal elliptical galaxies). \\

(b) The spectrum produced by low density gas ($n(H) \sim 200$\pcm) in pressure equilibrium with the X-ray cluster gas, when photoionised by the continuum of hot massive stars with $U$ in the range --3.5 to --4. As discussed, this reproduces the emission line ratios of the luminous `type II' systems, with [NII]/H$\alpha \simeq 0.6$, [SII]/H$\alpha \simeq 0.3$, [OIII]/H$\beta \simeq 0.75$, and again no appreciable \H2 emission. \\

(c) A population of dense ($n(H) \approxgt 10^{4}-10^{6}$\pcm), perhaps self-gravitating, clouds which are exposed to the same photoionising {\em flux} as those in (b); these produce the \H2 lines, with $\Sigma H_{\rm{2}}$/H$\alpha \sim 0.03-0.50$, and very little flux in the forbidden lines of [NII], [SII] and [OIII]. \\

Suppose that the fractions of the H$\alpha$ luminosity contributed by components (a), (b) and (c) are $f_{\rm{a}}$, $f_{\rm{b}}$ and $f_{\rm{c}}$, respectively, and that $f_{\rm{b}}=f_{\rm{c}}=0.5(1-f_{\rm{a}})$. Now if $f_{\rm{a}}$ decreases with H$\alpha$ luminosity from essentially unity at log L(H$\alpha$)=40 to around 0.2 at log L(H$\alpha$)=42, with a functional dependence chosen to yield the variation of [NII]/H$\alpha$ with log L(H$\alpha$) in Fig.~8 of Crawford et al.~(1999), then this also reproduces the variations of the [SII]/H$\alpha$, [OIII]/H$\beta$ and $\Sigma H_{\rm{2}}$/H$\alpha$ ratios with luminosity. In particular, the absence of \H2 emission in the lower luminosity objects (log L(H$\alpha$)$<41$) is accounted for. This phenomenological result, in which two or more components of emission appear to be relatively evenly-balanced over a range of more than two orders of magnitude in H$\alpha$ luminosity, perhaps implies that some sort of feedback is operating. However, the model parameter space and the scatter in the observed correlations are sufficiently large that it is difficult to say anything more quantitative. The most important result to emerge from the CLOUDY modelling is that a separate population of dense clouds is required to produce the \H2 emission. In the next subsection we discuss the origin of these clouds. 

Including dust in the CLOUDY models by means of the default `grains' option increases the total \H2~emission in the dense (component c) clouds by 20-30 per cent; the intrinsic optical emission produced by the clouds is affected much less. Although the extinction from the illuminated cloud surface to the point where the model integration stops (the point where the temperature has fallen to 200\K), is negligible ($A_{\rm{V}} \sim 10^{-3}$mag), predicting the observed spectrum is not straightforward. This is because grains at greater depth within the molecular cloud can scatter photons back with a wavelength-dependent albedo: the result is highly geometry-dependent. It should be emphasised that the gas-phase abundances were left unchanged at their solar values whilst the grains were introduced, so this illustrates that the grains {\em per se} do not have much effect. Using instead the `abundances ism' option in CLOUDY (which assumes depletion onto grains in accordance with observations), results in a reduction in the intrinsic \H2/H$\alpha$ ratio by a factor of $\sim 6$, owing to the existence of a much less extensive zone of warm ($>1000$K) \H2~in this case. The elemental abundances in the cold clouds in cooling flows may, however, be quite different from those in the ISM; emission line modelling (see Allen ~1995 and references therein) suggests metallicities of around solar; furthermore, recent {\em XMM-Newton} grating spectra of several clusters show a lack of emission lines from gas cooling below 1--2 keV, one explanation for which is an inhomogeneous distribution of metals within the ICM (Fabian et al.~2001).

To estimate the v=1-0~S(1) flux produced by the model for comparison with CLOUDY's $\Sigma H_{\rm{2}}$, we first extract the HI and \H2 densities and the temperature as functions of depth. For the gas densities of interest, we can assume collisional excitation of the \H2 and use the approximations described by Hollenbach \& Shull~(1979) for computing the fractions of \H2 in the various excited levels. Firstly, the fractions of molecules in each of v=0, 1 and 2 are estimated by approximating the molecule as a three level system. Assuming LTE for the rotational levels within v=1, and an ortho:para ratio of 3:1, the fraction of molecules in the v=1, J=3 state and hence the v=1-0 S(1) flux follow. For the above models we find v=1-0~S(1)/$\Sigma H_{\rm{2}}=0.22$, close to the figure estimated from the observed line fluxes listed in paper 1. To reproduce the 1-0 S(1)/H$\alpha$ ratios in Fig.~4, we thus need component (c) clouds with $\Sigma H_{\rm{2}}$/H$\alpha$ ratios (and hence densities), towards the upper end of the range considered above.

Using the same formalism, we can also calculate the populations of the v=0 rotational levels (which follow LTE up to some value $J=J_{\rm{max}}$, and then decrease more rapidly at higher J as the radiative decay rates increase), and hence the fluxes of the pure rotational 0-0 S series lines. For the component (c) clouds above, we find: 0-0~S(3)/1-0~S(1)$=0.4 \pm 0.1$, 0-0~S(2)/1-0~S(1)$\simeq 0.06$ and 0-0~S(1)/1-0~S(1)$\simeq 0.04$ (for our targets these are the best-placed lines, lying in the mid-infrared N and Q-bands). Although these 0-0 S line fluxes are around 10 times higher than predicted from an LTE extrapolation of the 1-0 S series line fluxes (using their observed excitation temperatures of $\sim 2000$\K), they will still challenge the capabilities of forthcoming mid-infrared spectrographs, such as Michelle on the UKIRT and Gemini telescopes. A measurement of the 0-0~S(3)/1-0~S(1) ratio would, however, provide a useful constraint on the variation of temperature with column density over the region of the cloud where the transition from atomic to molecular hydrogen occurs. 

%The interpretation of the CO observations of Edge~(2001), in the context of illuminated molecular cloud models, is addressed by Ferland, Fabian \& Johnstone~(2001).

\subsection{The physical picture}
One interpretation is that these dense gas clouds cooled directly from X-ray-emitting temperatures, as posited in the cooling flow scenario (see Ferland, Fabian \& Johnstone~1994 for a detailed discussion of the physical conditions within such clouds, albeit those at larger radii in the flow). If the high pressures in the \H2 gas are indeed due to self-gravitation, this is no surprise: self-gravitating molecular cloud cores must be abundant in copiously star-forming environments such as these. In this picture, the gas which produces the forbidden optical lines (component b above), is that which `boils off' these self-gravitating clouds, as proposed by Jaffe et al.~(2001).

The total mass of warm \H2 in these clouds, assuming it to be in LTE at 2000\K, is $4.2 \times 10^{4} L(S1)_{\rm{40}}$\Msun~where $L(S1)_{\rm{40}}$ is the luminosity of the v=1-0~S(1) line in units of $10^{40}$\ergps~(following Scoville et al.~1982). For the luminosities of interest this equates to masses of $10^{5-6}$\Msun. The gravitational collapse timescale of the clouds is $t_{\rm{grav}}=1/\sqrt{G\rho}=3 \times 10^{5} / \sqrt{n_{\rm{5}}}$\yr, where $n_{\rm{5}}$ is the particle number density in units of $10^{5}$\pcm. Thus, the warm molecular gas needs to be replenised at the rate of a few solar masses per year, a rate which is within an order of magnitude of the likely X-ray mass cooling rate in the central regions and the star-formation rate therein. 

As an alternative to self-gravity for producing the dense, high pressure clouds, one could appeal to shocks. Whilst there are considerable difficulties -- discussed in detail by Jaffe et al.~(2001) -- in constructing shock models which can {\em themselves} (i.e. through the shock emission) produce the observed \H2/H$\alpha$ ratios and the pressure differences in a manner consistent with the observed kinematics, one could instead invoke shocks just to create the dense, high pressure clouds. This could happen as follows. Suppose that two of the low density clouds of component (b) above, collide at a relative speed of a few hundred \kmps~(i.e. equal to the observed velocity dispersion in the CCG). The shocks are strong so the density jumps by a factor of 4 through the shock to $\sim 10^{3}$\pcm, and the immediate post-shock temperature is $10^{5}-10^{7}$\K; the pressure in the post-shock region is then $10^{8}-10^{10}$\pcmK, comparable to that of the \H2-emitting region. The gas then radiatively cools and its density simultaneously increases to maintain the high pressure. If it remains in the isobaric post-shock region until its temperature has dropped to $\sim 2000$\K, it will then come into equilibrium with the ambient radiation field of the young stellar population and emit the observed \H2 spectrum, exactly as in the case where the clouds are self-gravitating. The viability of this model depends on two conditions being satisfied: (i) on the shock geometry being such that the gas remains within the high pressure region until it has cooled to $\sim 2000$\K; (ii) on the line emission produced by the shocks {\em themselves} being negligible in comparison with that due to photoionisation of both the unshocked low density clouds and the shocked dense clouds. 

It is not clear whether either condition can be met. Firstly, if the clouds are magnetised, this will limit the post:preshock compression ratio to a maximum of $n_{\rm{m}}/n_{\rm{0}}=77v_{\rm{100}}/b$, where $v_{\rm{100}}100$\kmps~is the shock velocity and the magnetic field is assumed to scale with the preshock density as $B_{\rm{0}}=bn_{\rm{0}}^{0.5} \rm{\mu G}$ and $b$ is of order unity (see Hollenbach \& McKee~1979 for the derivation of this result). Once the gas has reached this density, any further cooling will be isochoric, not isobaric, so from a pre-shock density $n_{\rm{0}}=200$\pcm~the required densities of $10^{5}$\pcm~or more will not be attained unless the shock speed is at least 650\kmps. Secondly, at the speeds of interest the postshock material will itself be a powerful emitter of UV radiation and optical line emission: according to Dopita \& Sutherland~(1995), the UV and Balmer line fluxes emitted by the shock front are $8.9 \times 10^{-6} v_{\rm{100}}^{3.04} n_{\rm{0}}$ and $7.4 \times 10^{-6}v_{\rm{100}}^{2.41} n_{\rm{0}}$\ergpcmsqps, respectively. These figures are comparable to the external stellar flux incident on, and the resulting H$\beta$ line flux from, the dense post-shock gas in the CLOUDY model. There may thus be no need to invoke the existence of an additional stellar component: the radiation emitted by the shocked gas (which, for these shock speeds has a temperature of several times $10^{4}$\K- comparable to the temperatures of the massive stars invoked above) could serve the same purpose. This process, in which the UV radiation field and the dense gas are created together, could constitute the feedback referred to in section 2.3. Concerning the \H2~emission from the shocks themselves, we refer to the work by Hollenbach \& McKee~(1989) who simulated fast shocks (30-150\kmps), in gas at densities $n_{\rm{0}}=10^{3}-10^{6}$\pcm: such shocks are likely to be dissociative J-shocks from which the \H2 emission is due to formation pumping with a spectrum which resembles that of UV fluorescence. For $n_{\rm{0}} \approxgt 10^{5}$\pcm, there is a contribution from collisionally excited \H2~in gas at $\sim 500$\K, but this contribution is only significant for the pure rotational lines (0-0)S(0-5). The v=1-0~S(1)/H$\alpha$ ratios from such shocks lie below the observed range in Fig.~4 for all but the slowest shocks ($v < 50$\kmps).

Shocks have previously been invoked to account for the optical line emission in these systems. Crawford \& Fabian~(1992) used mixing layers to account for the spectra of the `type I' systems (component a above), combined with an increasing contribution of line emission from shocks in dense gas ($n_{\rm{0}} > 10^{5}$\pcm) to explain the transition to the more H$\alpha$-luminous `type II' systems. The latter authors also found correlations between the velocity dispersion and the [NII]/H$\alpha$ ratio, and between the local velocity dispersion and the H$\alpha$ surface brightness. Both correlations are naturally accounted for by the shock model, as discussed by Dopita \& Sutherland~(1995).

Further insight into the shock model, and an estimate of the ratio of the mass emitting H$\alpha$ to \H2, may be obtained as follows. Suppose that within a region of radius $R$ there exists a mean number density ${\cal N}$ of clouds of radius $r$, velocity $v$ and collisional cross-section $\sigma \sim \pi r^2$. The rate of collisions per unit volume is ${\cal N}^2 \sigma v$, so the mass of shocked gas at high pressure, given that the high pressure region lasts for a time $t \sim r/v$, is $M_{\rm p} \simeq \frac{4}{3} \pi R^3 n m_{p} \frac{4}{3} \pi r^3 {\cal N}^2 \sigma r$ (where $n$ is the particle density in the unshocked clouds, which we take to be $\sim 200$\pcm, as found for the component (b) clouds, and $m_{p}$ is the proton mass). The total mass of clouds is $M_{\rm c} \simeq \frac{4}{3} \pi R^3 {\cal N} n m_{p} \frac{4}{3} \pi r^3$. Thus $M_{\rm p}/M_{\rm c} = {\cal N} \pi r^3$. We may replace $M_{\rm p}$ by the mass of warm \H2~which is given above as $4.2 \times 10^{4} L(S1)_{\rm{40}}$\Msun; similarly, we may set $M_{\rm c}$ equal to the mass in the H$\alpha$ clouds, which is given by $1.2 \times 10^{5} L(H\alpha)_{\rm{40}}$\Msun~(using the expression for the H$\alpha$ emissivity for case B recombination at $10^{4}$\K~in Osterbrock~1989). Thus $M_{\rm p}/M_{\rm c} = 0.35 L(S1)/L(H\alpha)=0.0035-0.035$, since Fig.~4 shows that L(S1)/L(H$\alpha$) generally lies in the range 0.01--0.1 for $L(H\alpha)>10^{41}$\ergps. Thus the filling factor of the clouds, $f=\frac{4}{3} \pi r^3 {\cal N}$, lies in the range 0.005--0.05. The fact that the amount of shocked gas $\propto {\cal N}^2$ might explain why the \H2~emission is peaked towards the centre of the galaxy where ${\cal N}$ is likely to be highest.

It is also worth noting that the shocked gas may become self-gravitating. The post-shock cooling time $t_{\rm cool} \leq 10^{4}$\yr~(with equality holding for $10^{7}$\K), so $t_{\rm cool} \ll t_{\rm cross}$ unless $r < 1$\pc. The Jeans mass of the postshock gas is $M_{\rm J} = \frac{12 k^2 T^2}{\mu^2 G^{1.5} p^{0.5}} \sim 6 \times 10^{6} T_{4}^2 \mu_{-24}^{-2} p_{8}^{-0.5}$\Msun, where $\mu_{-24}$ is the mean mass per particle in units of $10^{-24}$\g. It is thus possible that collisions between the larger clouds ($r \approxgt 10$\pc, say) can produce self-gravitating objects with masses comparable to those of globular clusters, and may thus be relevant to the formation of the latter. Indeed, the specific frequency of globulars is sometimes high in cD galaxies (see e.g. the review of Harris~1991) and a young population has been seen at the centre of NGC 1275 (Holtzman et al.~1992). 

Since the shocked gas can become unstable to gravitational collapse and hence form stars, the gas cycling processes are undoubtedly complex. The above distinction between the shock and self-gravity scenarios for producing the dense gas may thus be too simplistic: some of the low density clouds may form by evaporation from the self-gravitating clouds, as well as by direct cooling out of the hot ICM, and both stars and UV radiation from shocked gas may contribute to the excitation.

%\subsection{Energetics}
%Using emission line ratios, we have establised that X-ray heating by the ICM is the most probable \H2 excitation mechanism. We now draw upon X-ray observations of cooling flow clusters to examine whether this is energetically feasible. On such grounds, Donahue et al. declared that it was remotely possible for PKS~0745-191 but highly unlikely for A2597. Such comparisons utilise the fact that for gas heated by X-rays to $\sim 2000$\K, about 0.5 per cent of the absorbed X-ray luminosity emerges as v=1-0~S(1); the figure of interest is thus the X-ray luminosity absorbed on spatial scales comparable to those probed by our spectra, which is $\sim 1 \times 1$~arcsec, since in many of the objects the \H2 emission is spatially-unresolved. To make such a comparison requires data with the spatial resolution of {\em Chandra}, although there may be some scope for extrapolation from {\em ROSAT} HRI deprojection studies.

\section{CONCLUSIONS AND FUTURE PROSPECTS}
Our UKIRT spectra demonstrate that the central galaxies of cooling flow clusters with H$\alpha$ luminosities above $10^{41}$\ergps~exhibit rovibrational \H2 line emission at the level of 0.01--0.1 H$\alpha$. The relative strengths of the \H2 lines imply that the emission is thermally excited in dense gas ($n \approxgt 10^{5}$\pcm) at temperatures $\sim 2000$\K, and is thus overpressurised by 2--3 orders of magnitude with respect to the optical emission line gas and the X-ray ICM. The emission could originate in a population of dense clouds heated by the young stellar populations which are known to exist in such systems and which can reproduce the optical forbidden line spectrum when incident on lower density gas. These dense clouds may be self-gravitating or confined in the high-pressure regions behind strong shocks. The fact that the high and low density clouds ultimately have a similar origin and are excited by the same radiation field accounts for the similar morphologies of the \H2 and optical emission line images found by Donahue et al.

Future observations at higher spectral resolution could address the kinematics of the \H2 emission for comparison with those of the CO emission (most of whose line widths are known to within 50\kmps~and lie in the range 100--300\kmps; Edge~2001), which may originate at a greater depth within the same dense clouds. Deeper spectra of an enlarged sample of low H$\alpha$ luminosity systems ($<10^{41}$\ergps) should also be obtained to establish at what level they emit \H2. Fig.~4 tentatively suggests that any \H2 in these systems may be produced by a different mechanism. Our models also open the possibility of detecting some of the purely rotational \H2 lines in the mid-infrared with forthcoming instruments such as Michelle on the UKIRT and Gemini telescopes. Observations of the 0-0~S(3)/1-0~S(1) ratio, for example, would provide a useful constraint on the variation of temperature with depth over the HI--\H2 transition region within the clouds.

\section*{ACKNOWLEDGMENTS}
UKIRT is operated by the Joint Astronomy Centre on behalf of the United Kingdom
Particle Physics and Astronomy Research Council. RJW acknowledges support from PPARC and an EU Marie Curie Fellowship. RMJ thanks PPARC, and ACE, CSC, ACF and SWA the Royal Society for support.

{}


\begin{thebibliography}{}

%\bibitem []{} Allen~S.W., Fabian~A.C., Johnstone~R.M., White~D.A., 
%Daines~S.J., Edge~A.C., Stewart~G.C., 1993, MNRAS, 262, 901
\bibitem []{} Allen~S.W., 1995, MNRAS, 276, 947
%\bibitem []{} Allen~S.W., Fabian~A.C., Edge~A.C., Bohringer~H., White~D.A, 
%1995, MNRAS, 275, 741
%\bibitem []{} Allen~S.W., Fabian~A.C., 1997, MNRAS, 286, 583
%\bibitem []{} Antonucci~R., Hurt~T., Kinney~A., 1994, Nature, 371, 313
%\bibitem []{} Barvainis~R., Antonucci~R., 1994, AJ, 107, 1291
%\bibitem []{} Barthel~P.D., K.A.~Arnaud, 1996, MNRAS, 283, L45
%\bibitem []{} Black J.H., 1998, Faraday Discussions, 109, 257, The Faraday Division of the Royal Society of Chemistry, London.
\bibitem []{} Black J.H., van Dishoeck E.F., 1987, ApJ, 322, 412
\bibitem []{} Brand P.J.W.L., Toner M.P., Geballe T.R., Webster A.S., Williams P.M., Burton M.G., 1989, MNRAS, 236, 929
%\bibitem []{} Conway~J.E., Blanco~P.R., 1995, ApJ, 449, L131
\bibitem []{} Crawford~C.S., Fabian~A.C., 1992, 259, 265
\bibitem []{} Crawford~C.S., Allen~S.W., Ebeling~H., Edge~A.C., Fabian~A.C., 
1999, MNRAS, 306, 857
\bibitem []{} Donahue M., Mack J., Voit G.M., Sparks W., Elston R., Maloney P.R., 2000, ApJ, 545, 670
\bibitem []{} Dopita M.A., Sutherland R.S., 1995, ApJ, 455, 468
%\bibitem []{} Draine~B.T., 1978, ApJS, 36, 595
%\bibitem []{} Draine~B.T., Roberge~W.G., Dalgarno~A., 1983, ApJ, 264, 485
%\bibitem []{} Draine B.T., Woods~D.T., 1990, ApJ, 363, 464
\bibitem []{} Edge~A.C., 2001, MNRAS, 328, 762
%\bibitem []{} Edge~A.C., Ivison~R.J., Smail~I., Blain~A.W., Kneib~J.-P., 1999, MNRAS, 306, 599
\bibitem []{} Edge~A.C., Wilman~R.J., Johnstone~R.M., Fabian~A.C., Allen~S.W., Crawford~C.S., 2002, to appear in MNRAS (paper 1)
%\bibitem []{} Fabian~A.C., 1994, ARAA, 32, 277
\bibitem []{} Fabian~A.C., Mushotzky~R.F., Nulsen~P.E.J., Peterson~J.R., 2001, MNRAS, 321, L20
\bibitem []{} Falcke~H., Rieke~M.J., Rieke~G.H., Simpson~C., Wilson~A.S., 1998, ApJ, 494, L155
\bibitem []{} Ferland~G.J., 1996, Hazy, A Brief Introduction to Cloudy, University of Kentucky Dept. of Physics and Astronomy Internal Report 
\bibitem []{} Ferland~G.J., Fabian~A.C., Johnstone~R.M., 1994, MNRAS, 266, 399
%\bibitem []{} Ferland~G.J., Fabian~A.C., Johnstone~R.M., 2001, submitted to MNRAS
%\bibitem []{} Ferguson~J.W., Korista~K.T., Ferland~G.J., 1997, ApJS, 110, 287
%\bibitem []{} Forbes~D.A., Ward~M.J., 1993, ApJ, 416, 150
\bibitem []{} Harris~W.E., 1991, ARA\&A, 29, 543
\bibitem []{} Hollenbach D., McKee C.F., 1979, ApJS, 41, 555
\bibitem []{} Hollenbach D., McKee C.F., 1989, ApJ, 342, 306
\bibitem []{} Holtzman~J.A., et al., 1992, AJ, 103, 6921
%\bibitem []{} Imanishi~M., Ueno~S., 2000, ApJ, 535, 626
%\bibitem []{} Jackson~N., Tadhunter~C., Sparks~W.B., Miley~G.K., Macchetto~F., 1996, A\&A, 307, L29
%\bibitem []{} Jackson~N., Tadhunter~C., Sparks~W.B., 1998, MNRAS, 301, 131
\bibitem []{} Jaffe~W., Bremer~M.N., 1997, MNRAS, 284, L1
\bibitem []{} Jaffe~W., Bremer~M.N., van der Werf~P.P., 2001, MNRAS, 324, 443
\bibitem []{} Krabbe~A., Sams~B.J., Genzel~R., Thatte~N., Prada~F., 2000, A\&A, 354, 439
%\bibitem []{} Landini~M., Natta~A., Salinari~P., Oliva~E., Moorwood~A.F.M., 1984, A\&A, 134, 284
\bibitem []{} Lepp S., McCray R., 1983, ApJ, 269, 560
\bibitem []{} Maloney~P.R., Hollenbach~D.J., Tielens~A.G.G.M., 1996, ApJ, 466, 561 
%\bibitem []{} Maloney~P.R., 1996, in {\em Cygnus A: Study of a Radio Galaxy}, eds. D.~Harris \& C.~Carilli, Cambridge University Press, Cambridge.
\bibitem []{} Mouri~H., 1994, ApJ, 427, 777
%\bibitem []{} Nussbaumer H., Storey~P.J., 1988, A\&A, 193, 327
%\bibitem []{} Ogle~P.M., Cohen~M.H., Miller~J.S., Tran~H.S., Fosbury~R.A.E., Goodrich~R.W., 1997, ApJ, 482, L37
%\bibitem []{} Osterbrock~D.E., Miller~J.S., 1975, ApJ, 197, 535
\bibitem []{} Osterbrock~D.E., 1989, Astrophysics of Gaseous Nebulae and Active Galactic Nuclei, University Science Books, Mill Valley, CA
%\bibitem []{} Peres~C.B., Fabian~A.C., Edge~A.C., Allen~S.W., Johnstone~R.M., White~D.A., 1998, MNRAS, 298, 416
%\bibitem []{} Reynolds~C.S., Fabian~A.C., 1996, MNRAS, 278, 479
\bibitem []{} Scoville~N.Z., Hall~D.N.B., Ridgway~S.T., Kleinmann~S.G., 1982, ApJ, 253, 136
%\bibitem []{} Shaw~M., Tadhunter~C.N., 1994, MNRAS, 267, 589
%\bibitem []{} Simpson~C., Forbes~D.A., Baker~A.C., Ward~M.J., 1996, MNRAS, 283, 777
\bibitem []{} Sternberg~A., Dalgarno~A., 1989, ApJ, 338, 197 (SD89)
%\bibitem []{} Stockton~A., Ridgway~S.E., Lilly~S.J., 1994, AJ, 108, 414
%\bibitem []{} Tadhunter~C.N., Scarrott~S.M., Rolph~C.D., 1990, MNRAS, 236, 163 
%\bibitem []{} Tadhunter~C.N., 1991, MNRAS, 251, L46
%\bibitem []{} Tadhunter~C.N., Metz~S., Robinson~A., 1994, MNRAS, 268, 989
%\bibitem []{} Tennant~A.F., 1991, NASA Technical Memorandum 4301
%\bibitem []{} Thornton~R.J., Stockton~A., Ridgway~S., 1999, AJ, 118, 1461
%\bibitem []{} Ueno~S., Koyama~K., Nishida~M, Yamauchi~S., Ward~M.J., 1994, ApJ, 431, 1
\bibitem []{} Voit~G.M., Donahue~M., 1997, ApJ, 486, 242
%\bibitem []{} Ward~M.J., Blanco~P.R., Wilson~A.S., Nishida~M., 1991, ApJ, 382, 115
%\bibitem []{} White~D.A., Fabian~A.C., Johnstone~R.M., Mushotzsky~R.F., Arnaud~K.A., 1991, MNRAS, 252, 72
%\bibitem []{} Whittle M., 1992, ApJ, 387, 109
\bibitem []{} Wilman R.J., Edge A.C., Johnstone R.M., Crawford C.S., Fabian A.C., 2000, MNRAS, 318, 1232
\end{thebibliography}
\end{document}